\documentclass[aps,prc,reprint,showpacs,showkeys,floatfix]
{revtex4-1}

\usepackage[dvips]{graphicx}
\usepackage{color}
\usepackage{subfigure}
\usepackage{dcolumn}     

\newcommand{\vi}[1]{\mbox{\boldmath $#1$}}
\newcommand{\vis}[1]{\mbox{\boldmath ${\scriptstyle #1}$}}
\newcommand{\mib}[1]{\mbox{\boldmath$#1$}}

\begin{document}

\title{Polarized proton+$^{4,6,8}$He 
elastic scattering with breakup effects in the eikonal approximation}
\author{K. Kaki}
\email[]{spkkaki@ipc.shizuoka.ac.jp}
\affiliation{Department of Physics, Shizuoka University, Shizuoka 422-8529, Japan}
\author{Y. Suzuki}
\affiliation{Department of Physics,  Niigata University, Niigata 950-2181, Japan}
\affiliation{RIKEN Nishina Center, Wakko 351-0198, Japan}
\author{R. B. Wiringa}
\affiliation{Physics Division, Argonne National Laboratory, Argonne, Illinois 60439, USA}

\date{\today}

\begin{abstract}
We study the elastic scattering of polarized protons from He isotopes. 
The central and spin-orbit parts of the optical potential are derived using   
the Glauber theory that can naturally take account of the 
breakup effect of the He isotopes. Both the differential cross section 
and the vector analyzing power for $p$+$^{4,6,8}$He scattering  
at 71\,MeV are in reasonable agreement with experiment. 
Scattering observables at 300\,MeV are predicted. 
The Pauli blocking effect is examined at 71\,MeV.  
\end{abstract}


\maketitle

\section{Introduction}

The spin-orbit potential for a nucleon moving in a nucleus
plays a decisive role in the nuclear shell structure. Though the 
phenomenological aspect of the spin-orbit potential has been 
clarified rather well for stable nuclei, a full account of its origin 
is still under active study. Important information on the spin-orbit 
potential is obtained through the analysis of elastic 
scattering observables, 
especially the vector analyzing power in elastic scattering of polarized protons
from the nucleus.

Recent developments of experimental techniques have provided 
us with a polarized proton target and have made it
possible to measure not only the differential
cross section but also the vector analyzing power for elastic scattering of
polarized protons from unstable nuclei in inverse kinematics. 
Data have recently been taken for $^6$He~\cite{tu10} and $^8$He~\cite{ss08} at
71~MeV/nucleon. A good example of a two-neutron halo
nucleus is $^6$He, where the neutron density extends far out in distance. This
unique feature of its structure is expected to show up in the vector 
analyzing power because the spin-orbit potential is primarily 
sensitive to the surface 
of the nucleus.  Though less 
pronounced than $^6$He, $^8$He has an extended neutron cloud as well and its 
elastic scattering observables are interesting. 

There have been so far only a few theoretical studies on the optical 
potential and 
observables for $p$+$^{6,8}$He elastic scattering. The single scattering
approximation to the multiple scattering expansion was employed in the
$p$+$^8$He case~\cite{crespo}. The predicted angular distribution of the 
vector analyzing power shows a peak 
at about $\theta_{\rm c.m.}$=46$^{\circ}$, which is different from 
the value obtained from the data~\cite{ss08}. The 
$p$+$^{8}$He elastic scattering angular distribution was analyzed 
in the eikonal model to examine its sensitivity to the 
matter distribution of $^8$He~\cite{chulkov}. The  $p$+$^{6,8}$He 
elastic scattering observables were calculated in a full-folding optical
model~\cite{spw00}. The calculated vector analyzing powers do not 
agree with experiment~\cite{tu10,ss08}. Very recently 
neutron pickup coupling with the elastic channel has been studied 
to see its effect on proton scattering from $^6$He~\cite{km11}. 
As noted above, none of the calculations succeeds in reproducing the 
vector analyzing powers for $p$+$^{6,8}$He scattering. It is reported
that a search for the  optical potential parameters of 
the spin-orbit part leads to a shallow and long-ranged 
spin-orbit potential for $^6$He~\cite{tu10}. 

The purpose of this paper is to analyze the elastic scattering 
observables, the differential cross 
section, the vector analyzing power and the spin-rotation function, 
for protons scattered from He isotopes including $^4$He~\cite{sb89}. 
The central part of the 
optical potential for the proton is calculated 
in the framework of the Glauber or eikonal model~\cite{glauber}. The 
inputs needed in the calculation include only the ground state wave
functions of the He isotopes and the nucleon-nucleon scattering 
amplitude, or more precisely its Fourier transform, the nucleon-nucleon
profile function. 
The spin-orbit potential is constructed by using a derivative of the 
central part of the optical potential. There are at least three 
noticeable advantages of the 
present approach: First, it is logically very simple, and nevertheless 
it contains nucleon-nucleon multiple scatterings to all orders. Second,
the wave function of the projectile nucleus itself can be 
employed, though 
in an approximate version of the present approach 
the projectile density may be used instead of the wave function. 
Third the optical potential obtained takes account of 
breakup effects of the projectile without recourse to 
laborious calculations with continuum
discretization~\cite{yabana,suzuki03}, which enables us to 
discuss easily the dynamic polarization potential (DPP). 
The breakup effect is a vital ingredient 
that should be taken care of for the optical potential of a 
weakly bound nucleus such as $^6$He. 

In Sec.~\ref{formulation} we present a formulation needed in 
our approach. We show in Sec.~\ref{op} how to construct the 
optical potential, its approximate version and 
a relationship between those potentials and a folding potential.
We define our spin-orbit 
potential in Sec.~\ref{observable} together with the elastic 
scattering observables.  In Sec.~\ref{vmcwf} we outline 
the variational Monte Carlo 
method that is employed to generate the ground state wave functions of the He 
isotopes.  
Results of our calculation are presented in Sec.~\ref{result}. First we 
compare the elastic scattering observables at 71~MeV/nucleon with
experiment in Sec.~\ref{analysis.71}. The 
angular distributions of the differential cross section and vector
analyzing power for $p$+$^4$He scattering at higher energies 
are compared to available data in Sec.~\ref{analysis.high}. 
Predictions for $p$+$^{6,8}$He
scattering at intermediate energy are also given. 
Conclusions are drawn in Sec.~\ref{conclusion}.

\section{Formulation}
\label{formulation}
\subsection{Optical potential in the eikonal approximation}
\label{op}

Let us assume that the projectile nucleus moves in
the $z$ direction and impinges on a proton target with a velocity $v$. 
Let ${\vi R}$ stand 
for the relative distance vector between the projectile and the proton. 
Under the eikonal approximation the $x, y$ component of ${\vi R}$, 
denoted ${\vi b}$, turns out to be just a parameter, the
impact parameter. The interaction $v_{pN}$ between 
the $i$th nucleon of the 
projectile and the proton gives rise to a multiplicative 
phase factor 
$e^{i\chi_{pN}({\vis b}-{\vis s}_i)}$~\cite{glauber} that modifies the wave function of the
projectile, where ${\vi s}_i$ is the $x, y$ component of the position 
vector ${\vi r}_i$ 
of the $i$th nucleon relative to the center of mass of the
projectile. The phase
$\chi_{pN}$ is related to $v_{pN}$ by 
\begin{equation}
\chi_{pN}({\vi b})=-\frac{1}{\hbar v} \int_{-\infty}^{+\infty}
v_{pN}(\sqrt{{\vi b}^2+z^2})dz.
\label{chi.V}
\end{equation}
Each nucleon contributes a position-dependent 
multiplicative phase factor. A
key quantity to describe the proton-projectile elastic scattering is
given by the eikonal phase
\begin{eqnarray}
{\rm e}^{i\chi_{\rm E}({\vis b})}=\langle\Phi_0 |\, {\rm e}^{i\Xi({\vis b}, {\vis s}_1,
\ldots, {\vis s}_A)}| \Phi_0 \rangle,
\label{psf}
\end{eqnarray}
where $\Xi({\vi b}, {\vi s}_1,\ldots, {\vi s}_A)$=
$\sum_{i=1}^A\chi_{pN}({\vi b}-{\vi s}_i)$ is the total phase, and 
$\Phi_0$ is the ground state wave function of the projectile nucleus.  

As is well known~\cite{yabana,suzuki03}, the optical phase shift
function~(\ref{psf}) obtained in the eikonal approximation includes 
the effects of coupling with excited states or breakup continuum
states. To make this point clear, we define the average of the total
phase 
\begin{equation}
\chi_{\rm F}({\vi b})=\langle \Phi_0 |\, \Xi({\vi b}, {\vi s}_1,\ldots, {\vi s}_A)| \Phi_0 \rangle.
\label{folding.psf}
\end{equation}
Hereafter the projectile nucleus is assumed to be 
spherical, so that both $\chi_{\rm E}(\vi b)$ and 
$\chi_{\rm F}(\vi b)$ become a function of $b=|\vi b|$.
Using Eqs.~(\ref{chi.V}) and (\ref{folding.psf}), we find that
$\chi_{\rm F}(b)$ is the phase shift function corresponding to the
single folding potential 
$U_{\rm f}(R)$=$\int \rho_N(r)v_{pN}(|{\vi R}-{\vi r}|)d{\vi r}$
\begin{equation}
\chi_{\rm F}(b)=-\frac{1}{\hbar v}\int_{-\infty}^{+\infty}
U_{\rm f}(\sqrt{{\vi b}^2+z^2}) dz,
\end{equation}
where $\rho_N(r)$ is the nucleon density of the
projectile nucleus. The eikonal phase $\chi_{\rm E}(b)$ is expressed as

\begin{equation}
\chi_{\rm E}(b)=\chi_{\rm F}(b)-i{\rm ln}
\langle \Phi_0 |\, {\rm e}^{i(\Xi({\vis b}, {\vis s}_1,\ldots, {\vis
s}_A)-\chi_{\rm F}(b))}| \Phi_0 \rangle.
\label{eikonalphase}
\end{equation}
According to Glauber~\cite{glauber}, a potential 
\begin{eqnarray}
U_{\rm c}(R) 
     & = & \frac{\hbar v}{\pi} \frac{1}{R} \frac{d}{dR} 
           \int_0^{\infty}  \chi_{\rm E}( \sqrt{R^2+x^2} )dx
\label{cpot}
\end{eqnarray}
produces the phase shift function $\chi_{\rm E}(b)$
in the eikonal approximation.
The potential $U_{\rm c}(R)$ differs from $U_{\rm f}(R)$ in that the
underlying phase shift function of the former contains the second term
of the right-hand side of Eq.~(\ref{eikonalphase}). The term can be
discussed in a cumulant expansion~\cite{glauber,suzuki03} 
that involves the fluctuation of the 
higher order cumulants $\langle \Phi_0 |\,(\Xi({\vi b}, {\vi s}_1,\ldots, {\vi
s}_A)-\chi_{\rm F}(b))^n| \Phi_0 \rangle$. 
The difference between 
$U_{\rm c}(R)$ and the folding potential $U_{\rm f}(R)$ is 
the DPP. As is clear from the above derivation, 
the DPP is evaluated in the 
eikonal approximation without an explicit invoking of the couplings with 
excited and continuum states.  

It should be stressed that
the central part of the optical potential $U_{\rm c}(R)$ can be obtained
in a unified way independently of the projectile nucleus.

The above formulation has successfully been applied to a study of the 
breakup effects of weakly bound projectile nuclei, e.g., $^2$H scattered
from $^{58}$Ni~\cite{yabana} and $^6$He scattered from 
$^{12}$C~\cite{abu-ibrahim03}. In these applications the
target is not a proton but a composite nucleus. It is treated as an 
absorbing point particle and then it is possible to apply exactly the
same formulation as above by adopting appropriate nucleon-target optical
potentials for $v_{pN}$. The quality of such calculations is tested by
comparing to other calculations that explicitly include 
the breakup channels in the continuum discretized coupled-channels 
method~\cite{yahiro,matsumoto}. 

For the proton target, $v_{pN}$ stands for the proton-nucleon potential. 
Any operator dependence of the potential has to be avoided because 
otherwise the evaluation of Eq.~(\ref{psf}) together with Eq.~(\ref{chi.V}) 
is impossible. Instead of looking for some effective forces 
that have no operator dependence, we
follow a simple procedure here.  We introduce 
the proton-nucleon profile function $\Gamma_{pN}(\vi b)$, which is 
equal to $1-{\rm e}^{i\chi_{pN}({\vis b})}$. With 
$\Gamma_{pN}(\vi b)$, 
the proton-nucleon scattering amplitude $f_{pN}(\theta)$ is given in 
the eikonal approximation by 
\begin{equation}
f_{pN}(\theta)=\frac{iK}{2\pi}\int {\rm e}^{-i{\vis q}\cdot{\vis b}}\,
 \Gamma_{pN}(\vi b)\, d{\vi b},
\end{equation}
where $K$ is the wave number and ${\vi q}$ is the momentum transfer,
$q$=$2K\sin \frac{\theta}{2}$, and
$\Gamma_{pN}(\vi b)$ is conveniently parametrized  as 
\begin{eqnarray}
\Gamma_{pN}(\mib b) = \frac{1-i \alpha_{pN}}{4 \pi \beta_{pN}} \
 \sigma_{pN}^{\rm tot} \, {\rm e}^{-\frac{{\vis b}^2}{2 \beta_{pN}}},
\label{profile}
\end{eqnarray}
where $\alpha_{pN}$ is the ratio of the real to the imaginary part of the {\it pN}
scattering amplitude in the forward direction, 
$\beta_{pN}$ is the slope parameter of the {\it pN} elastic differential cross section,
and $\sigma_{pN}^{\rm tot}$ is the {\it pN} total cross section 
due to the nuclear {\it pN} interaction. 
We use the parameter values tabulated in Ref.~\cite{abi08}. 
The difference between $pp$ and $pn$ interactions is taken
into account in what follows by extending Eq.~(\ref{profile}) to 
\begin{equation}
\Gamma_{pN}(\mib b)=\delta_{Np}\Gamma_{pp}(\mib b)+\delta_{Nn}\Gamma_{pn}(\mib b).
\end{equation}
In this way we bypass the direct use of the nuclear force in calculating 
$\chi_{\rm E}(b)$. 

The eikonal phase $\chi_{\rm E}(b)$ and the average total phase
$\chi_{\rm F}(b)$ are expressed in terms of the profile function as
follows
\begin{eqnarray}
\label{psf.profile}
&&\hspace{-5mm}\chi_{\rm E}(b)=-i\log \langle\Phi_0 |\, \prod_{i=1}^A\big(1-\Gamma_{pN}(\vi
 b-\vi s_i)\big)| \Phi_0 \rangle, \\
&&\hspace{-5mm}\chi_{\rm F}(b)=-i\int \rho_N(r)\log \Big(1-\Gamma_{pN}(\vi
 b-\vi s)\Big)\,d{\vi r}.
\label{psf.folding}
\end{eqnarray}
Both neutron and proton densities are employed in calculating $\chi_{\rm
F}(b)$. As will be explained 
in Sec.~\ref{vmcwf}, $\chi_{\rm E}(b)$ is 
obtained with a Monte Carlo integration. It turns out that
the present approach together with the Monte Carlo integration is very 
versatile for calculating the eikonal phase and the corresponding
optical potential. A simpler calculation is to
take the leading order of the cumulant expansion, 
leading to
\begin{equation}
\chi_{\rm E}^{(1)}(b)=i\int \rho_N(r)\Gamma_{pN}(\vi b-\vi s)\,d\vi r.
\label{cumulant}
\end{equation}
This provides us with a reasonable approximation to the full phase 
for the nucleon-nucleus scattering, and is often adopted for the 
evaluation of the phase shift function for the scattering between
complex
nuclei~\cite{abi00,wh07,abi08}. Equation~(\ref{cpot}) is
used to obtain those potentials which generate the phases $\chi_{\rm
F}(b)$ and $\chi_{\rm E}^{(1)}(b)$, and they are compared  
with $U_{\rm c}(R)$.

\subsection{Elastic scattering observables}
\label{observable}
The potential $U_{\rm c}(R)$ constructed in this way is central
and includes the effect of breakup of the projectile
in the eikonal approximation. 
The spin-orbit term of the optical potential is introduced as
\begin{eqnarray}
U_{\rm so}(R) = V_{\rm so} \lambda_{\pi}^2 
\frac{1}{R} \frac{d}{dR} U_{\rm c}(R),
\label{lspot}
\end{eqnarray}
where $\lambda_{\pi}$ is the pion Compton wavelength. 
The real and imaginary parts of the 
spin-orbit potential are thus obtained from the real and imaginary parts of 
$U_{\rm c}$, respectively 
and the constant $V_{\rm so}$ that determines the spin-orbit
strength is allowed to be
different depending on the real and imaginary parts of the spin-orbit 
potential.

Proton-nucleus scattering is described in 
a partial wave expansion 
with the potential $U_{\rm c}(R)+\mib l \cdot \mib \sigma
U_{\rm so}(R)$.
The scattering amplitude becomes an operator in the spin space
\begin{eqnarray}
\hat{f}(\theta) = f(\theta) + i g(\theta) \mib \sigma \cdot \hat{\mib n}
\end{eqnarray}
with a unit vector $\hat {\mib n}$ perpendicular to 
the scattering plane
\begin{eqnarray}
\hat {\mib n} = \frac{ \mib k \times \mib k'}{| \mib k \times \mib k' |},
\end{eqnarray}
where $\mib k$ and $\mib k'$ are the momenta in the center
of mass before and after the scattering.
Denoting the $S$ matrix by $S_l^{\pm} = \exp( 2i \delta_l^{\pm} )$ 
where $\delta_l^{\pm}$ are complex phase shifts for the potential
$U_{\rm c}({R}) + l U_{\rm so}({R})$ or 
$U_{\rm c}({R}) - (l+1)U_{\rm so}({R})$, respectively, 
we obtain
\begin{widetext}
\begin{eqnarray}
f(\theta) & = & f_C(\theta) + \frac{1}{2ik} \sum_l 
               \left[(l+1)(S_l^+ -1) + l(S_l^- -1) \right] \ e^{2i \delta_l^C}
               P_l(\cos \theta), \\
g(\theta) & = & \frac{1}{2ik} \ \sum_l (S_l^+ - S_l^-)\ e^{2i \delta_l^C} P_l^1(\cos \theta),
\end{eqnarray}
\end{widetext}
where $f_C(\theta)$ is the Coulomb scattering amplitude, 
$\delta_l^C$ is the Coulomb
phase shift, and $P_l^1(\theta)$ is the associated Legendre
polynomial. A deviation of the Coulomb potential from that of a point
charge is included in the calculation but its effect is very small.

For a fixed angle $\theta$ the scattering amplitude $\hat f$ is determined by three
real quantities.
They can be conventionally chosen as the differential cross section
$d \sigma/d \Omega$, the vector analyzing power $A_y$, and the
spin-rotation function $Q$: 
\begin{eqnarray}
\frac{d \sigma}{d \Omega} & = & | f(\theta)|^2 + | g(\theta)|^2, \\ 
A_y(\theta)               & = & \frac{2 {\rm Re} (f(\theta) g^*(\theta) )}
                                     {| f(\theta)|^2 + | g(\theta)|^2}, \\
Q(\theta)                 & = & \frac{2 {\rm Im} (f(\theta) g^*(\theta) )}
                                     {| f(\theta)|^2 + | g(\theta)|^2}.
\end{eqnarray}

\subsection{Variational Monte Carlo wave function}
\label{vmcwf}

The wave functions of He isotopes used in this work are taken from
variational Monte Carlo (VMC) calculations for a Hamiltonian consisting of
nonrelativistic nucleon kinetic energy, the Argonne $v_{18}$ two-nucleon 
potential~\cite{WSS95}, and the Urbana IX three-nucleon potential~\cite{PPCW95}:
\begin{eqnarray}
H = { \sum_{i} K_i } + { { \sum_{i<j}} v_{ij} }
+ { \sum_{i<j<k} V_{ijk} } \ .
\end{eqnarray}
A VMC calculation finds an upper bound $E_V$ to an eigenenergy $E_0$ of the 
Hamiltonian by evaluating the expectation value of 
$H$ in a trial wave function, $\Psi_V$:
\begin{eqnarray}
E_V = \frac{\langle \Psi_V | H | \Psi_V \rangle}
{\langle \Psi_V | \Psi_V \rangle} \geq E_0  \ .
\end{eqnarray}
The parameters in $\Psi_V$ are varied to minimize $E_V$, and the lowest value 
is taken as the approximate energy.
The multidimensional integral is evaluated using standard Metropolis
Monte Carlo techniques~\cite{MR2T2} (and hence the VMC designation).
A good trial function is given by~\cite{WPCP00}
\begin{eqnarray}
   |\Psi_V\rangle = {\cal S}\prod_{i<j}^A
      \left[1 + U_{ij} + \sum_{k\neq i,j}^{A}\tilde{U}_{ijk} \right]
                    |\Psi_J\rangle \ ,
\label{eq:psiv}
\end{eqnarray}
where $U_{ij}$ and $\tilde{U}_{ijk}$ are noncommuting two- and three-body 
correlation operators induced by the dominant parts of $v_{ij}$ and $V_{ijk}$, 
respectively; ${\cal S}$ is a symmetrizer, and the Jastrow wave function 
$\Psi_J$ is
\begin{eqnarray}
   |\Psi_J\rangle = \prod_{i<j}f_c(r_{ij}) |\Phi_A(J^\pi;T T_z)\rangle \ .
\end{eqnarray}
Here the single-particle $A$-body wave function $\Phi_A(J^\pi;T T_z)$ is fully 
antisymmetric and has the total spin, parity, and isospin quantum numbers 
of the state of interest, while the product over all pairs of the central 
two-body correlation $f_c(r_{ij})$ keeps nucleons apart to avoid the strong
short-range repulsion of the interaction.
The long-range behavior of $f_c$ and any single-particle radial dependence
in $\Phi_A$ (which, to ensure translational invariance, is written using 
coordinates relative to the center of mass of the $s$-shell core)
control the finite extent of the nucleus.
For $p$-shell nuclei, there are actually three different central pair
correlation functions $f_c$: $f_{ss}$, $f_{sp}$, and $f_{pp}$,
depending on whether both particles are in the $s$-shell core ($ss$), both
in the $p$-shell valence regime ($pp$), or one in each ($sp$).

The two-body correlation operator has the structure 
\begin{eqnarray}
   U_{ij} = \sum_{p=2,6} u_{p}(r_{ij}) O^{p}_{ij} \ ,
\end{eqnarray}
where the $O^{p}_{ij}$ are the leading spin, isospin, spin-isospin, tensor,
and tensor-isospin operators in $v_{ij}$.
The radial shapes of $f_c(r)$ and $u_p(r)$ are obtained by numerically 
solving a set of six Schr\"{o}dinger-like equations: 
two single-channel ones for $S$=0, $T$=0 or 1,
and two coupled-channel ones for $S$=1, $T$=0 or 1, with the latter producing
the important tensor correlations~\cite{W91}.
These equations contain the bare $v_{ij}$ and parametrized Lagrange 
multipliers to impose long-range boundary conditions of exponential decay
and tensor/central ratios.

Perturbation theory is used to motivate the three-body correlation operator
\begin{eqnarray}
\tilde{U}_{ijk} = -\epsilon \tilde{V}_{ijk}(\tilde{r}_{ij},
                     \tilde{r}_{jk}, \tilde{r}_{ki})
\end{eqnarray}
where $\tilde{r}=yr$, $y$ is a scaling parameter, $\epsilon$ is a (small
negative) strength parameter, and $\tilde{V}_{ijk}$ includes the dominant
short-range repulsion and anticommutator part of two-pion exchange in the 
three-nucleon potential.
Consequently, $\tilde{U}_{ijk}$ has the same spin, isospin, and tensor 
dependence that $\tilde{V}_{ijk}$ has.

The variational parameters in $f_{ss}$, $U_{ij}$, and $\tilde{U}_{ijk}$ have 
been chosen to minimize the energy of the $s$-shell nucleus $^4$He.
For the $p$-shell nuclei $^6$He and $^8$He, these parameters are kept
fixed and the additional parameters that enter $f_{sp}$, $f_{pp}$, and the 
single-particle radial behavior of $\Phi_A$ have been adjusted to minimize 
the energy of these systems subject to the constraint that the proton and 
neutron rms radii are close to those obtained from more sophisticated Green's 
function Monte Carlo (GFMC) calculations~\cite{PPCPW97,WPCP00}.

The wave function samples used here are generated by following a
random walk guided by the $\Psi_V$ for each nucleus.
After an initial randomization, a move is attempted, where each particle
is randomly shifted within a box of 1.2-1.4 fm in size; the $\Psi_V$ is
evaluated and compared to the previous configuration, with the move 
being accepted or rejected according to the Metropolis algorithm.
After ten attempted moves, the configuration is saved, including the 
$x,y,z$ coordinates of each particle (with the center of mass being set to zero)
and the probability for each particle to be either a neutron or a proton.
The size of the box gives an acceptance rate of $\sim$50\% and we generate
one million configurations for each nucleus.

\begin{table}
\caption{
Root-mean-square radii, given in units of femtometers,
of the proton and neutron distributions of $^{4,6,8}$He. 
The proton radius of the second column is obtained by converting 
the measured charge radius.
}
\label{tab.1}
\begin{ruledtabular}
\begin{tabular}{cccc} 
He isotope    & Experiment~\cite{pm07}    &
 \multicolumn{2}{c}{Calculated} \\
\cline{3-4}
              & proton    & proton  & neutron \\
\hline
$^{4}$He      & 1.457(4)  & 1.447   & 1.447 \\
$^{6}$He      & 1.938(23) & 1.928   & 2.871 \\
$^{8}$He      & 1.885(48) & 1.884   & 2.901 \\ 
\end{tabular}
\end{ruledtabular}
\end{table}

The root-mean-square radii calculated from the VMC wave 
functions are listed in Table~\ref{tab.1}. 
The second column shows the proton root-mean-square 
radii extracted from the charge
radii that are obtained from the experiment
based on laser spectroscopy~\cite{pm07}.
The results of the VMC wave functions 
excellently reproduce the radii determined 
experimentally.

\begin{figure}
\scalebox{0.4}{\includegraphics{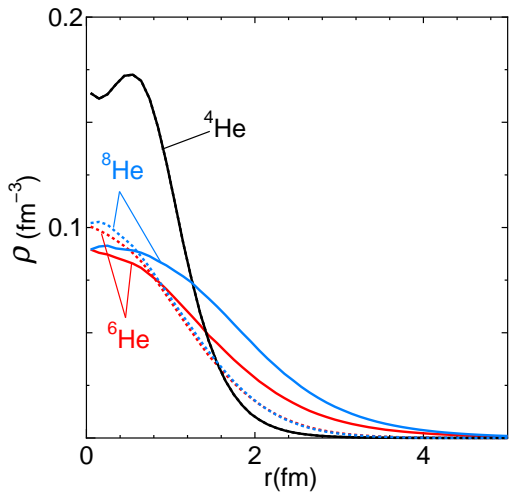}}
\caption{(Color online)
Density distributions of $^{4,6,8}$He calculated with the VMC wave functions.
Solid and dotted lines denote the neutron and proton
 distributions, respectively.}            
\label{fig.1}
\end{figure}

The density distributions of He isotopes 
are displayed in Fig.~\ref{fig.1}.
Solid and dotted lines denote the neutron and proton 
distributions, respectively.  Both proton and neutron 
distributions of $^4$He are almost the same and are confined to a small
region. Though the proton
distribution is very similar in both $^6$He and $^8$He, 
the tail of $^6$He
is slightly more extended to larger distances. Beyond 2\,fm, the neutron
distribution overwhelms that of the proton. The falloff of $^6$He neutron 
density is rather slow compared to that of $^8$He owing to its weak binding.

\section{Results}
\label{result}

The construction of the optical potential, (\ref{cpot}) and
(\ref{lspot}), requires the $3A$-dimensional integration indicated in 
Eq.~(\ref{psf}) or (\ref{psf.profile}). This integration can be performed with ease in the Monte Carlo
method using the wave function samples generated in Sec.~\ref{vmcwf}. 
The feasibility and the accuracy of the Monte Carlo integration in the
calculation of the phase shift function was already demonstrated for
various cases in Ref.~\cite{kv02}. The differentiation 
in Eq.~(\ref{cpot}) is facilitated by fitting $\chi_{\rm E}(b)$ 
in terms of several Gaussians with different falloff parameters:  
$\chi_{\rm E}(b)$=$\sum_k C_k \exp(-a_kb^2)$. 

\subsection{Scattering at 71\,MeV}
\label{analysis.71}
First we discuss 
the elastic scattering observables at 71\,MeV. 
The strength $V_{\rm so}$ of the spin-orbit potential is the
only parameter in the present formalism. The
differential cross section is largely determined by the central
potential $U_{\rm c}$. The spin-orbit potential contributes to the 
differential cross section almost negligibly within a reasonable range of
the strength, and so we try 
to fit the vector analyzing power by 
varying $V_{\rm so}$. In the case of $^4$He, we choose different values 
of the strength parameter depending on the spin-orbit real or imaginary
potential. 
The spin-orbit imaginary potential for $^6$He and $^8$He is 
set to be zero for the sake of simplicity.  
The depth parameters of the spin-orbit potential are listed in 
Table~\ref{tab.2}. The value of $V_{\rm so}$ for 
the $p$+$^4$He spin-orbit real potential is consistent with that employed 
in the systematics of one-particle motion~\cite{bohr}. Compared to
$^4$He, the strength of the spin-orbit real potential for $^{6,8}$He 
is considerably smaller, in accordance with the analysis of Ref.~\cite{tu10}.

\begin{table}
\caption{Strength parameters $V_{\rm so}$ of 
the spin-orbit potentials  
for $p$+$^{4,6,8}$He scattering at 71\,MeV.
}
\begin{ruledtabular}
\begin{tabular}{ccc} 
He isotope  & Real potential & Imaginary potential \\ \hline
$^4$He       & 0.125      & $-$0.075 \\
$^6$He       & 0.025      & \ \ 0.00 \\
$^8$He       & 0.05       & \ \ 0.00 \\ 
\end{tabular}
\end{ruledtabular}
\label{tab.2}
\end{table}

\begin{figure*}
\subfigure[~Central potentials]{
\scalebox{2.0}
{\includegraphics[clip,width=0.5\columnwidth]{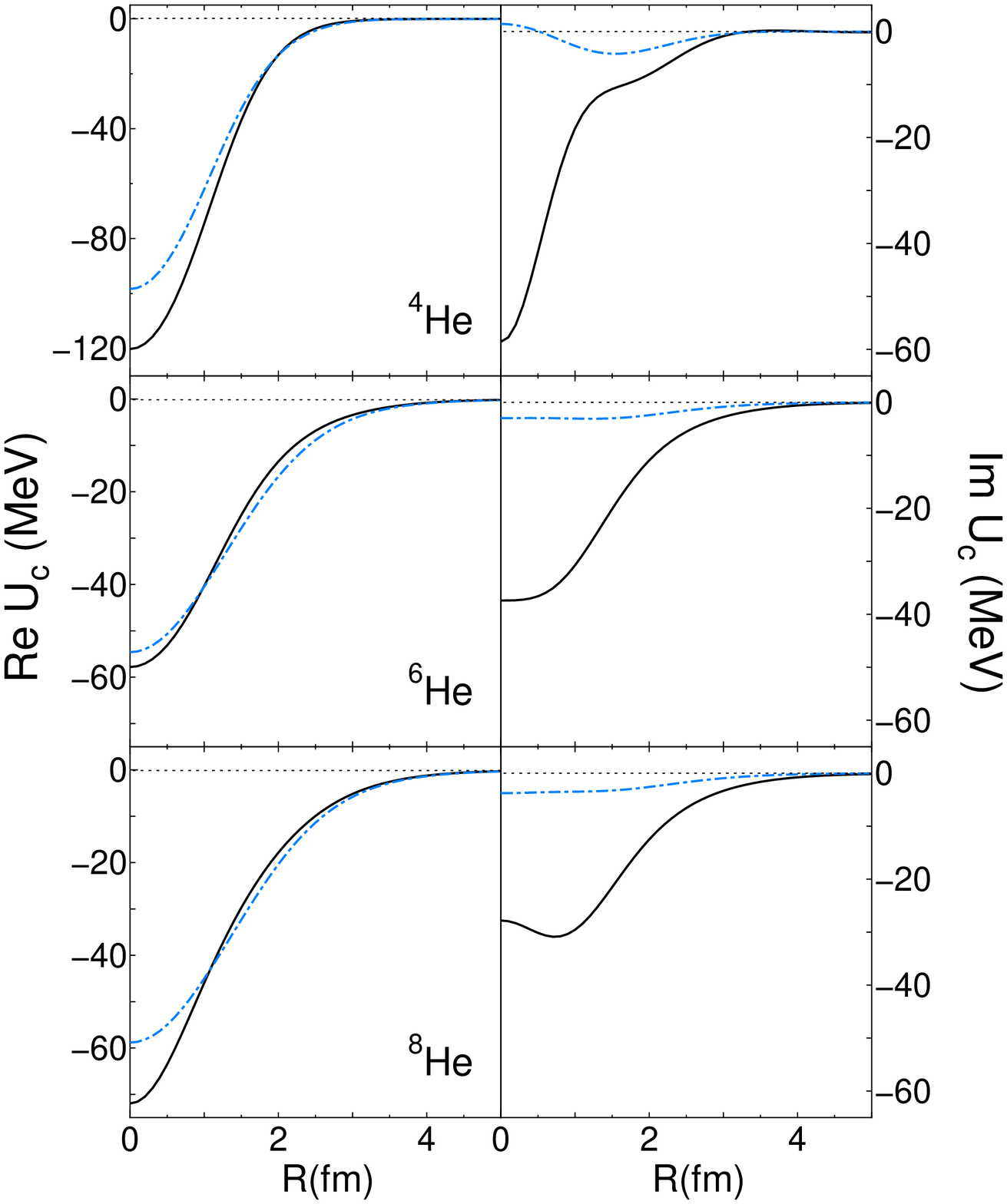}}
}
\subfigure[~Spin-orbit potentials]{
\scalebox{2.0}
{\includegraphics[clip,width=0.5\columnwidth]{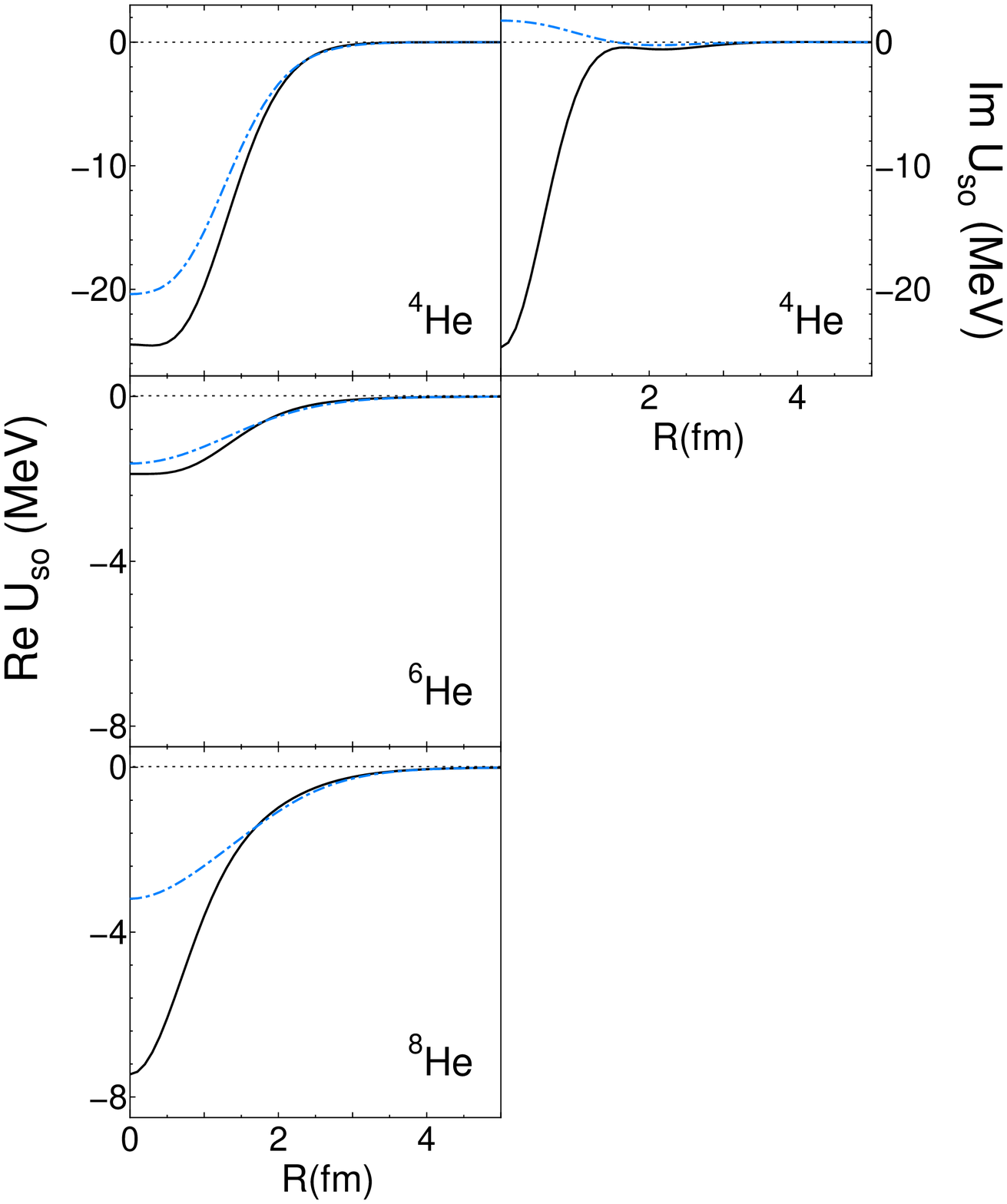}}
}
\caption{(Color online) 
Optical potentials for $p$+$^{4,6,8}$He 
elastic scattering at 71\,MeV. The central and spin-orbit potentials are 
shown in Fig.~2(a) and 2(b), respectively. 
The left and right panels are the real and imaginary potentials,
 respectively. 
The spin-orbit imaginary potentials of $^6$He and $^8$He 
are set to be zero.
Solid and dash-dotted lines are the potentials of the full eikonal model and 
the folding model, respectively.
}
\label{fig.2}
\end{figure*}

Figure~\ref{fig.2} shows the  
optical potentials calculated for $p$+$^{4,6,8}$He elastic scattering.  
The left and right panels of Fig.~2(a) are the 
central real and imaginary potentials. Solid lines denote potentials 
of the full calculation, while dash-dotted lines are the folding potentials. The
optical potential for $^4$He is much deeper than the others 
at short distances due to its compact structure 
but becomes much shallower near the surface. Compared to the folding
potential,  the full optical potential has the following property 
near the surface: the imaginary part turns out to
be much more absorptive and the real part is less attractive. This is a
general feature of the DPP due to the breakup effect 
as already observed in Ref.~\cite{yabana}. Similarly 
Fig.~2(b) exhibits the spin-orbit  real and imaginary 
potentials. 

\begin{figure*}
\scalebox{0.80}{\includegraphics{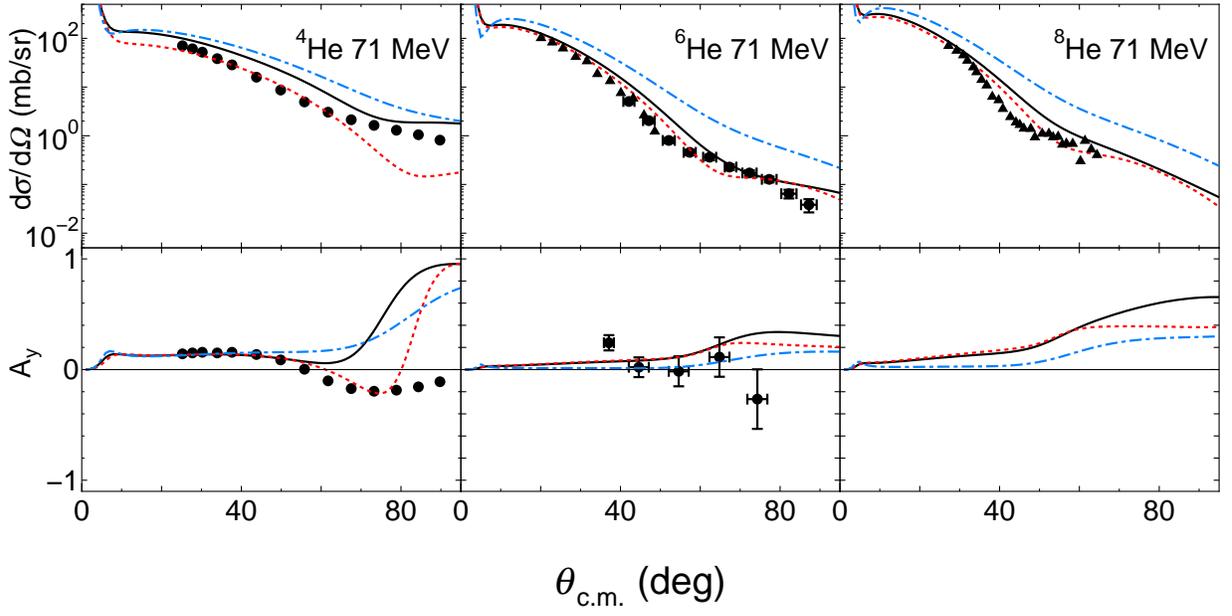}}
\caption{(Color online) Angular distributions for $p$+$^{4,6,8}$He 
elastic scattering at 71\,MeV. 
The upper panels show the differential cross sections
and the lower panels the vector analyzing powers. 
Solid lines are full eikonal calculations~(\ref{psf.profile}), 
dotted lines are the approximation in the leading-order cumulant 
expansion~(\ref{cumulant}), and 
dash-dotted lines are the folding  model calculations~(\ref{psf.folding}).  
Experimental data 
are taken from  Ref.~\cite{sb89} for $^4$He, 
from Refs.~\cite{tu10,korshen97} for $^6$He, and from
Refs.~\cite{korshen93} for $^8$He.}
\label{fig.3}
\end{figure*}

Figure~\ref{fig.3} displays the differential cross sections (upper panels) 
and vector analyzing powers (lower panels) for $^4$He, $^6$He 
and $^8$He calculated  using the above potentials. 
Solid lines denote the
results based on the potential derived from the full phase shift 
function~(\ref{psf.profile}), dotted lines are the results 
obtained from its 
leading order in the cumulant expansion~(\ref{cumulant}), and
dash-dotted lines are the results of the folding
potential~(\ref{psf.folding}). It is seen that the full potential 
gives smaller and better differential cross sections 
than the folding potential. The vector analyzing powers are also 
better reproduced with the full potential. The leading order
approximation seems to be surprisingly good. The 
vector analyzing power available for $^4$He is accurately 
known~\cite{sb89}, 
and its behavior that changes sign around 60$^{\circ}$ demands 
a nonvanishing $V_{\rm so}$ value for the imaginary potential.   
In Ref.~\cite{km11} the effects of a DPP due to 
pickup coupling on the differential cross section and 
vector analyzing power 
for proton elastic scattering from $^6$He has been studied. 
Their results for full the calculation including both pickup and breakup 
contributions 
have shown a decrease in the differential cross section around 50$^{\circ}$ 
and an increase in the vector analyzing power at angles larger than 
50$^{\circ}$, which is in disagreement with the data.

Though the present theory  reproduces the 
experimental data --- especially the vector analyzing powers ---
fairly well, the 
calculated differential cross sections tend to be  
large compared 
to experiment. The incident energy of 71\,MeV is likely not 
high enough for the Glauber approximation. In such a case 
that the incident energy of the projectile is comparable to its 
Fermi energy, the use of information on the free space  
nucleon-nucleon interaction may not be valid, but Pauli-blocking 
effects may become important. In fact the chance for  the incident 
nucleon to collide with the proton target will be suppressed 
because of the Pauli effect. 

\begin{figure*}
\scalebox{0.8}{\includegraphics{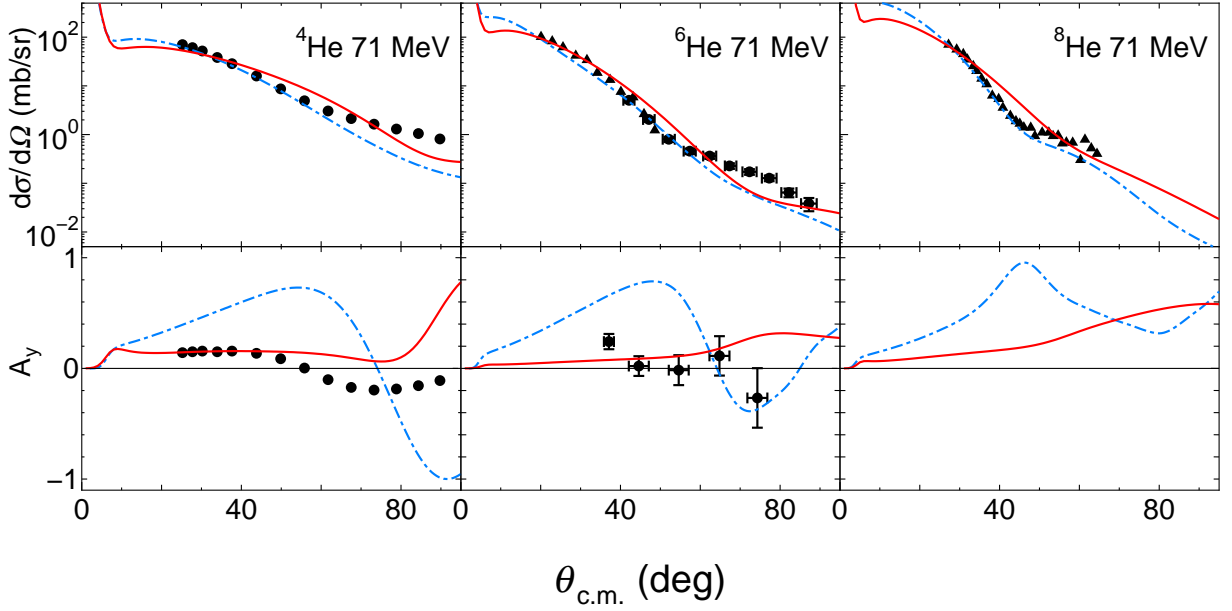}}
\caption{(Color online) Angular distributions for $p$+$^{4,6,8}$He 
elastic scattering at 71\,MeV compared between calculations and 
experiment. 
Solid lines are the results of the present model 
with the Pauli-blocking effect, while dash-dotted 
lines are the results 
taken from Ref.~\cite{spw00}. 
See the caption of Fig.~\ref{fig.3} for the experimental data.}
\label{fig.4}
\end{figure*}

\begin{figure}
\scalebox{0.45}{\includegraphics{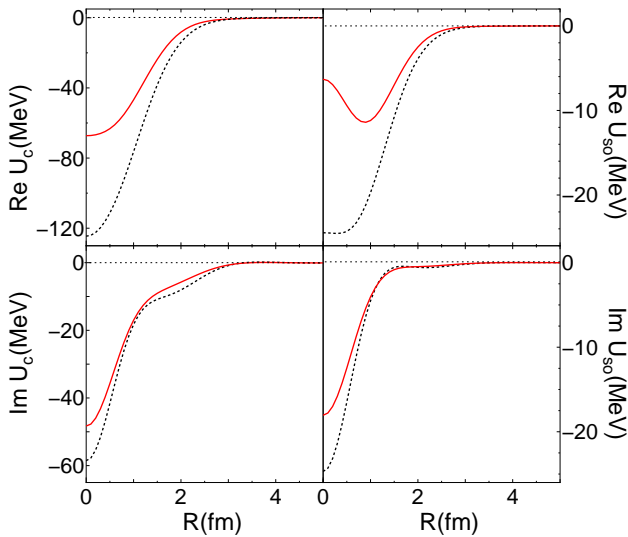}}
\caption{(Color online)~
Optical potentials for $p$+$^4$He elastic scattering at 
71\,MeV. 
The upper two panels show the central and spin-orbit real potentials
 while the lower two panels show the imaginary potentials. 
Solid and dotted lines denote the potentials of the full calculation 
with and without the Pauli-blocking effect, respectively.
}
\label{fig.5}
\end{figure}

To simulate this Pauli-blocking correction
phenomenologically, we replace the 
$\sigma_{pN}^{\rm tot}$ value of Eq.~(\ref{profile}) with
\begin{equation}
\bar{\sigma}_{pN}^{\rm tot}=\sigma_{pN}^{\rm tot}\left(1-\gamma \frac{7}{5}
\frac{E_F}{E}\right),
\label{paulicorrection}
\end{equation}
where  the Fermi energy $E_F$ is 
related, in the local density approximation, to the nucleon 
density $\rho_N(r)$ by  
$E_F=\frac{\hbar^2}{2m_N}(3\pi^2\rho_N(r))^{2/3}$.  
The value of $\gamma$ is unity according to 
Ref.~\cite{hussein91}, but 
here it is adjusted to reproduce the differential cross 
section at forward angles. 
In fact the use of
Eq.~(\ref{paulicorrection}) with $\gamma=1$
for $^4$He turns out to lead to unphysical 
cross sections because 
the $^4$He density is very large and 
changes drastically in a small region and thus 
the local density approximation may not work so well. 
The adopted value of $\gamma$ is 0.5 for all He 
isotopes. At the same time, 
the range parameter 
$\beta_{pN}$ is also appropriately changed as noted in 
Ref.~\cite{abi08}.  
The values of $V_{\rm so}$ are kept unchanged. 
The solid line in Fig.~\ref{fig.4} denotes the 
result of the calculation with the Pauli-blocking correction. 
Compared to the solid line in Fig.~\ref{fig.3} that has no 
correction, we see that the Pauli-blocking correction results in the decrease of the differential 
cross sections, leading to fair agreement with experiment at 
forward angles. The vector 
analyzing power for $^4$He is very much improved as well. 
Also plotted in Fig.~\ref{fig.4} are the results of 
Ref.~\cite{spw00} in which the optical potential based on the 
Watson formulation of multiple-scattering theory was derived. Our calculation 
including the breakup effect clearly offers a better description 
of the scattering of the proton with the He isotopes. 
The optical potential with the Pauli-blocking correction is displayed 
in Fig.~\ref{fig.5}, and it is compared to the one without the 
correction. Both real and imaginary parts are reduced significantly 
for $r < 2 $, fm by the Pauli-blocking effect.

As seen in Fig.~\ref{fig.4}, the theory slightly underestimates the $p+^4$He 
differential cross section beyond 70 degrees. We considered the 
effect of the knock-on process to see whether or not that gives an 
important contribution in improving the cross section. 
In the knock-on process the incident 
nucleon knocks a nucleon inside 
the nucleus and the knocked nucleon is ejected out of the nucleus. 
The knock-on exchange effect produces a nonlocal potential and 
it was calculated by assuming a $(0s)^4$ 
wave function of $^4$He and the Minnesota central potential~\cite{mn}. 
This nonlocal potential is transformed to an equivalent local 
potential following the WKB procedure~\cite{horiuchi}. The effect 
of this potential is, however, so small that the above discrepancy remains to 
be resolved.

\subsection{Scattering at 300 and 500\,MeV}
\label{analysis.high}

Next we study the elastic scattering observables at 300 and 500\,MeV 
where $p$+$^4$He data are available for both differential cross 
sections and  vector analyzing powers~\cite{my01,sms92}. 
The eikonal approximation should work better at these high energies. 

\begin{figure}
\scalebox{0.45}{\includegraphics{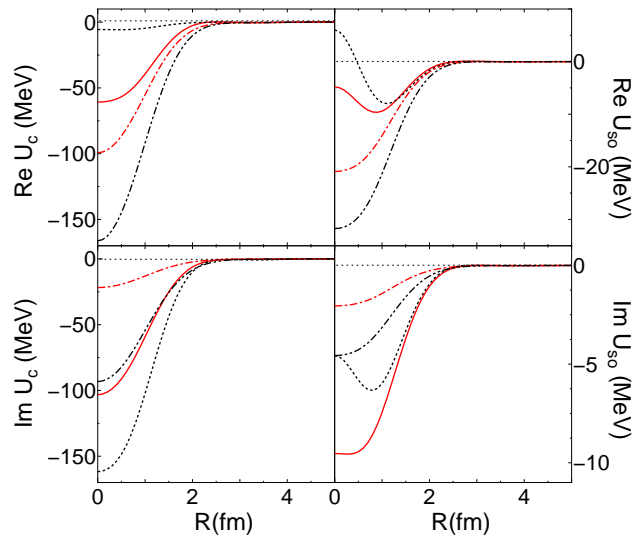}}
\caption{(Color online)
Optical potentials for $p$+$^4$He elastic scattering at 300 and
 550\,MeV. 
The upper two panels show the central and spin-orbit real potentials
 while 
the lower two panels show the imaginary potentials. 
Solid and dash-dotted lines denote the full and folding potentials 
at 300\,MeV, whereas dotted and dash-dot-dotted lines denote the full and folding 
potentials at 550\,MeV.
}
\label{fig.6}
\end{figure}

Figure~\ref{fig.6} displays the optical potentials for proton-elastic scattering 
from $^4$He at 300 and 550\,MeV.
As already known~\cite{abu-ibrahim03}, it is seen that with increasing projectile 
incident energy
the depth of the central real potential decreases and that of the central imaginary potential 
increases. 
The central real potential calculated in the full model 
at 550\,MeV is found to be extremely shallow, which results in a very small 
real part of the spin-orbit potential if $V_{\rm so}$ is of the order of 0.1. 
In order to account for the vector analyzing power, $V_{\rm so}$ has to be chosen 
to be about 1.0 as will be shown below.

\begin{figure}
\scalebox{0.5}{\includegraphics{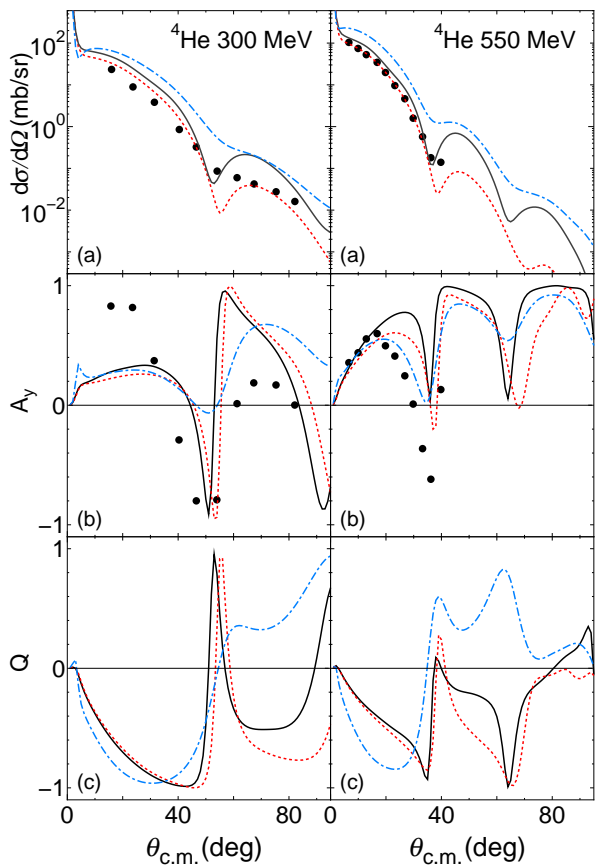}}
\caption{(Color online)  Angular distributions for $p$+$^{4}$He 
elastic scattering at 300 and 550\,MeV. Displayed are 
the differential cross section (a), the vector analyzing power
 (b), and the spin-rotation function (c). The values of 
$V_{\rm so}$ are 0.1 and $-$0.05 for the real and imaginary potentials at 
300\,MeV, and 0.9 (0.09 for the folding calculation) and $-$0.025 at 550\,MeV. 
Solid, dotted, and dash-dotted lines indicate the same 
types of calculations as those of Fig.~\ref{fig.3}.
Experimental data are taken from Ref.~\cite{my01} for 
300\,MeV and 
from Ref.~\cite{sms92} for 500\,MeV. 
}
\label{fig.7}
\end{figure}

Figure~\ref{fig.7} exhibits observables of proton elastic scattering 
from $^4$He at 300\,MeV (left side) and 550\,MeV (right side).
From the top, differential cross section (a), 
vector analyzing power (b), and spin-rotation function (c)
are given.
Dots are experimental data \cite{my01} for 300\,MeV and \cite{sms92} for 500\,MeV.
Note that the theory 
at 550\,MeV is compared to the experiment at 500\,MeV. The present theory 
reproduces both the differential cross section and the vector analyzing power 
reasonably well.

\begin{figure}
\scalebox{0.5}{\includegraphics{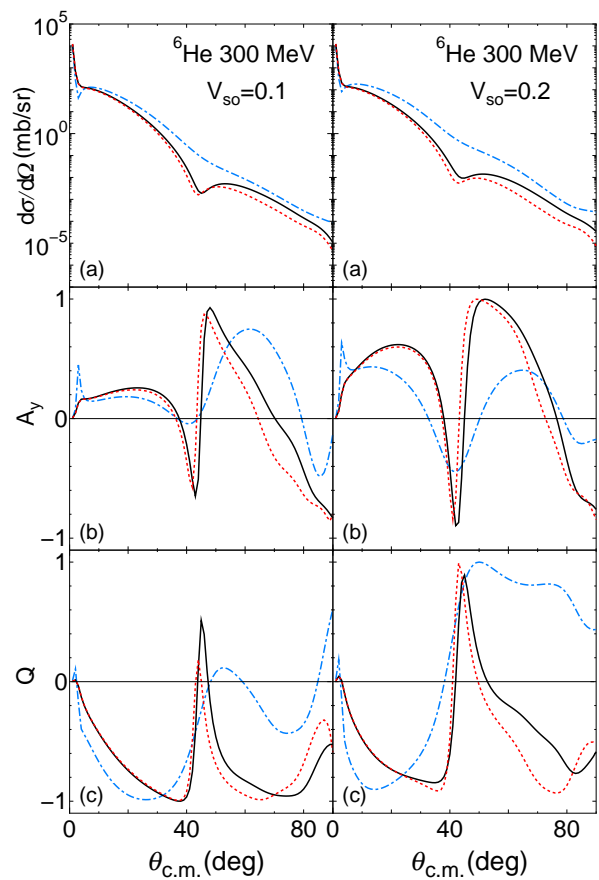}}
\caption{(Color online) Angular distributions for $p$+$^{6}$He 
elastic scattering at 300\,MeV. 
The left panels show the results calculated with $V_{\rm so}$=0.1 for the real potential,  
the same strength as 
that of $^4$He at 300\,MeV, while the right panels show the results with
 $V_{\rm so}$=0.2. The spin-orbit imaginary potential is set
 to be 0.05, which is the same as that of $^4$He. Solid, dotted, and
 dash-dotted lines indicate the same types of calculations as those of  
Fig.~\ref{fig.3}.
}
\label{fig.8}
\end{figure}

\begin{figure}
\scalebox{0.5}{\includegraphics{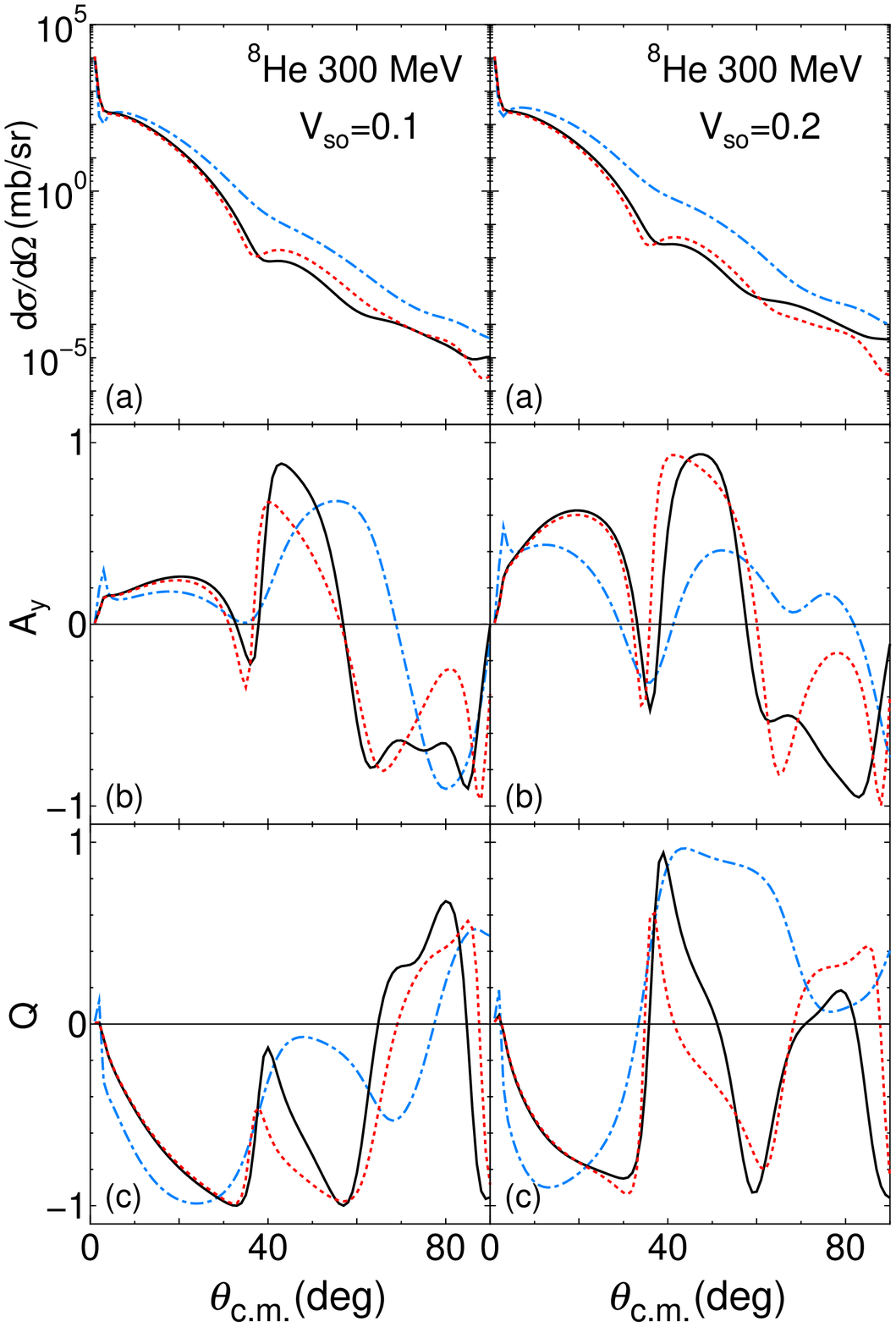}}
\caption{ The same as Fig.~\ref{fig.8} but for $p$+$^{8}$He 
elastic scattering at 300\,MeV. 
}
\label{fig.9}
\end{figure}

Finally we predict the scattering observables at 300\,MeV for  
$p+^6$He in Fig.~\ref{fig.8} and  for $p+^8$He in Fig.~\ref{fig.9}.
Solid and dotted lines in both figures show results for full and folding model calculations.  
In the case where the real depth parameter $V_{\rm so}$ 
is chosen to be the same as that of the $^4$He case at 300\,MeV, 
the angular distributions for $^6$He and $^8$He are 
similar to that for $^4$He at 300\,MeV, as expected. 
Calculations with a parameter twice as large 
($V_{\rm so}=0.2$) 
increase the vector analyzing powers at small angles and the differential cross sections at large angles. 
The differential cross section, the vector analyzing 
power, and the spin-rotation function are all different  
between the full and folding model calculations at angles larger than 50$^{\circ}$, 
as we observe in Figs.~\ref{fig.3} and \ref{fig.7}.

\section{Conclusion}
\label{conclusion}

We have analyzed the elastic scattering observables for 
protons scattered from He isotopes at 71 and 300\,MeV. 
The optical potentials for $p+^{4,6,8}$He systems are 
calculated in the Glauber model. The central potential 
is evaluated to all orders of the complete Glauber amplitude 
using the nucleon-nucleon scattering amplitude and the 
ground-state wave function of the He isotope that is 
taken from the variational Monte Carlo method. 
Both the real and imaginary parts of the central potential 
are determined without any adjustable parameters. It should  
be noted that the central potential obtained in this way 
takes into account the breakup effect of the He isotope 
to its excited states including continuum states, which  makes it possible to learn the difference from the single folding model potential. The spin-orbit potential is 
assumed to 
take the standard form that uses the derivative of the 
central potential. Its strength is the only parameter 
in the present approach. 

Though the incident energy of 71\,MeV may be a little too low for applying 
the Glauber theory, 
the present theory leads us to reasonable agreement with 
experimental data especially on the vector analyzing powers 
for $p$+$^{4,6,8}$He scatterings simultaneously. 
It should be noted here that usual $t$-folding calculations  fail to reproduce 
the vector analyzing powers.
We observe that the differential cross sections are 
all slightly larger than the experimental data even at 
forward angles. We have studied the Pauli-blocking 
effect that partly suppresses the interaction between the proton 
and the He isotope and found that including it 
improves the angular distribution of the elastic 
scattering.

For higher incident energy scattering from $^4$He, the differential cross section 
is reproduced well at forward angles and the 
vector analyzing power
is also reproduced well in accordance with the experimental data.
The vector analyzing power in the intermediate energies has been a long-standing problem 
that defies reproduction 
of the experimental data. 
However, the spin-orbit potential given by the derivative of the central potential
calculated from the Glauber theory appears to be able to 
reproduce the proton elastic scattering
in both the low- and intermediate-energy regions. We have 
predicted the angular distributions for $p$+$^{6,8}$He 
elastic scattering at 300\,MeV.

\section{Acknowledgments}
The authors thank T. Uesaka and S. Sakaguchi for several useful
discussions and Ch. Elster for sending them the results of 
Ref.~\cite{spw00}.  
The work of Y.~S. is supported in part by Grants-in-Aid for 
Scientific Research (No. 21540261 and No. 24540261) of the Japan Society for the 
Promotion of Science. 
The work of R.~B.~W. is supported by the U.~S.~Department of Energy, Office of Nuclear 
Physics, under Contract No. DE-AC02-06CH11357.

\end{document}